\documentclass{svproc}
\usepackage{url}
\usepackage{lineno}
\usepackage{graphicx}
\begin{document}
%\pdfoutput=1
\mainmatter              % start of a contribution
\title{Measurement of charged jet cross sections and jet shapes in proton-proton collisions at 
$\surd s $~=~2.76~TeV with the ALICE detector at LHC}
\titlerunning{Charged jet properties}  % abbreviated title (for running head)
\author{Rathijit Biswas for the ALICE Collaboration}
\authorrunning{Rathijit Biswas} % abbreviated author list (for running head)
\tocauthor{Rathijit Biswas}
\institute{Centre for Astroparticle Physics and Space Science, Bose Institute, Kolkata,~India\\
\email{rathijit.biswas@cern.ch}
}
%\linenumbers
\maketitle              % typeset the title of the contribution
\begin{abstract}
We present measurements of charged jet cross sections and jet shape observables in leading jet in proton-proton (pp) 
collisions at $\surd s $~=~2.76~TeV with the ALICE detector at LHC. Jets are reconstructed at midrapidity from charged
particles using the sequential recombination anti-k$_{T}$ jet finding algorithm for various R values. The results are compared 
to measurements from HERWIG, PHOJET and different tunes of PYTHIA6 and earlier measurements at 7~TeV.
\end{abstract}
\section{Introduction}
Jets are defined as the collimated stream of final state hadrons, produced from the fragmentation of hard scattered partons 
(quarks and gluons) in high energy collisions. Jets serve as tool to connect the final state hadrons with the primarily
produced partons and thus they are probe to test perturbative quantum chromodynamics (pQCD). The study of jet shapes provides
details of the parton to jet fragmentation with insight to jet tranverse profile. Furthermore, the measurements in pp 
serves as a baseline reference for similar measurements in heavy-ion (AA) collisions, where a highly dense medium is created,
to isolate the hot matter effects. 
\section{Data analysis, jet reconstruction and observables}
For this analysis, 4.4 X 10$^7$ minimum bias events recorded by the ALICE detector at the Large Hadron Collider (LHC) during 2011
are used. Events with primary vertex within $\pm 10$ cm along the beam axis from the geometrical interaction point are 
analysed. Charged tracks are reconstructed using the combined information from Time Projection Chamber (TPC)
and Inner Tracking System (ITS) and tracks with p$_{T}^{track}$ $>$ 150 MeV/c, $\mid$ $\eta^{track}$ $\mid$ $<$ 0.9 are
selected for the analysis.
Jets are reconstructed using sequential recombination anti-k$_{T}$ \cite{antikt} jet clustering algorithm from
the FastJet package \cite{fastjet}.
The differential charged jet cross sections are measured with R = 0.2, 0.3, 0.4 and 0.6 using the equation~ 
 ~\begin{equation}
  \frac{d^{2}\sigma ^{jet, ch}}{dp_{T}d\eta}(p_{T}^{jet, ch}) = \frac{1}{\mathcal{L}^{int}} \frac{\Delta N_{jets}}
  {\Delta p_{T}\Delta \eta}(p_{T}^{jet,ch})
 ~\end{equation}~
  ~where $\mathcal{L}$$^{int}$ is the integrated luminosity (1.3 nb$^{-1}$ in our case) and $\Delta$N$_{jets}$ the number of jets in the selected 
  intervals of $\Delta$p$_{T}$ and $\Delta$$\eta$. Jet shape
  observables such as average numbers of charged tracks in a 
  leading jet ($\langle$N$_{ch}\rangle$), leading charged jet size (average radius containing 80\% of the total jet 
  p$_{T}$, $\langle$R$_{80}\rangle$) and radial distribution of p$_{T}$ density about the jet axis (dp$_{T}$$^{sum}$/dr) are measured for
  R = 0.4.
\section{Corrections and systematic uncertainties}
The jet p$_{T}$ spectra are corrected for instrumental effects by the Bayesian unfolding technique \cite{unfold} using 
Monte Carlo (MC) events and full detector simulation with GEANT3~\cite{geant}. Jet shape observables are
corrected using a bin-by-bin correction technique. 
All the observables are further corrected for the contribution of the underlying event (UE) from 
sources other than hard scattering. The UE is estimated following the
technique used by ALICE~\cite{ue} and subtracted event-by-event
for jet p$_{T}$ spectra and on a statistical basis for jet shape observables.
The sources of systematic uncertainties include finite efficiency and momentum resolution, UE 
estimation and the dependence of correction factors on MC event generators.
\section{Results}
\begin{figure}[h]
  \centering
  \includegraphics*[scale = 0.10]{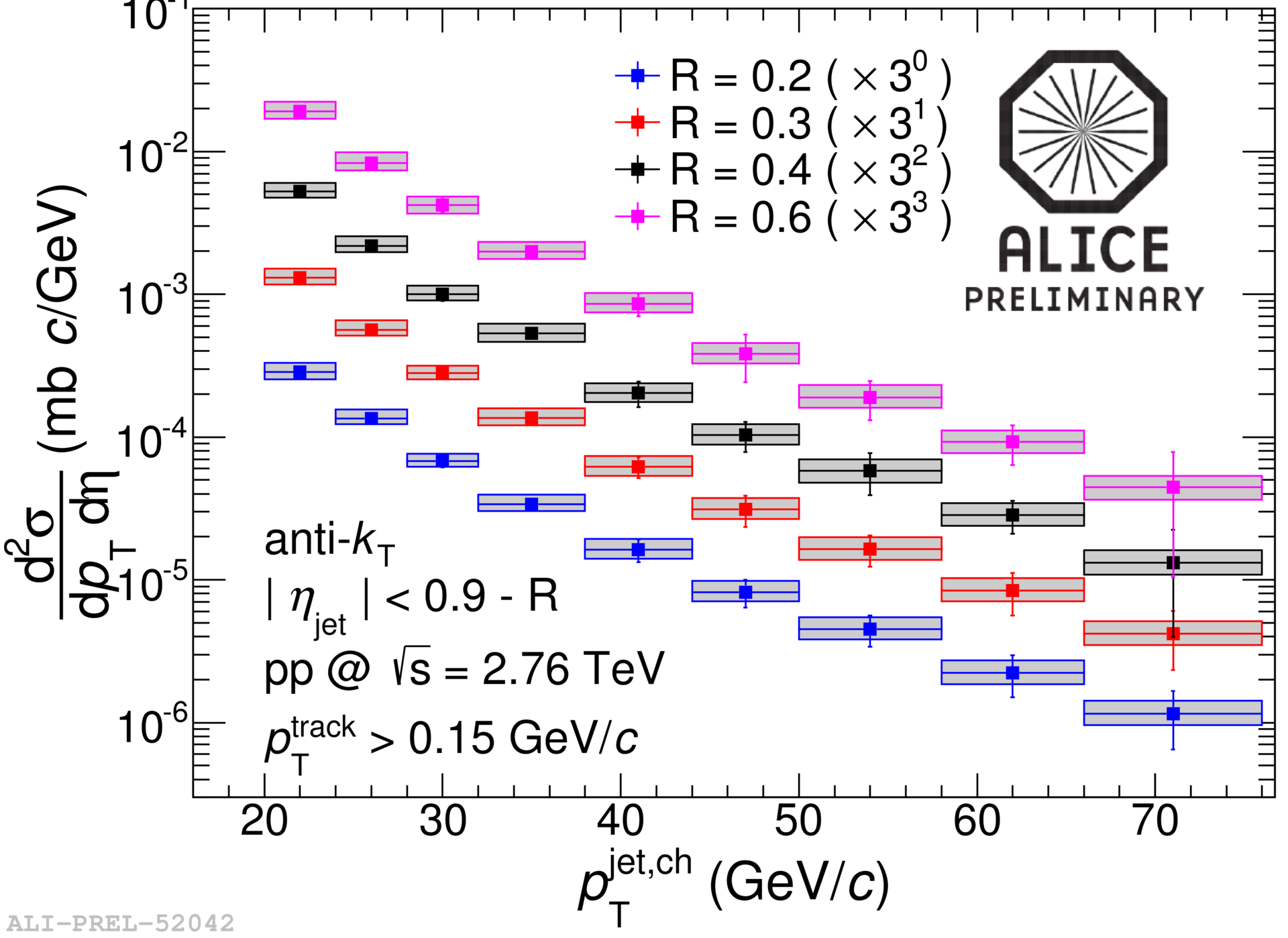}
   \includegraphics*[width = 6 cm, height = 5.7 cm]{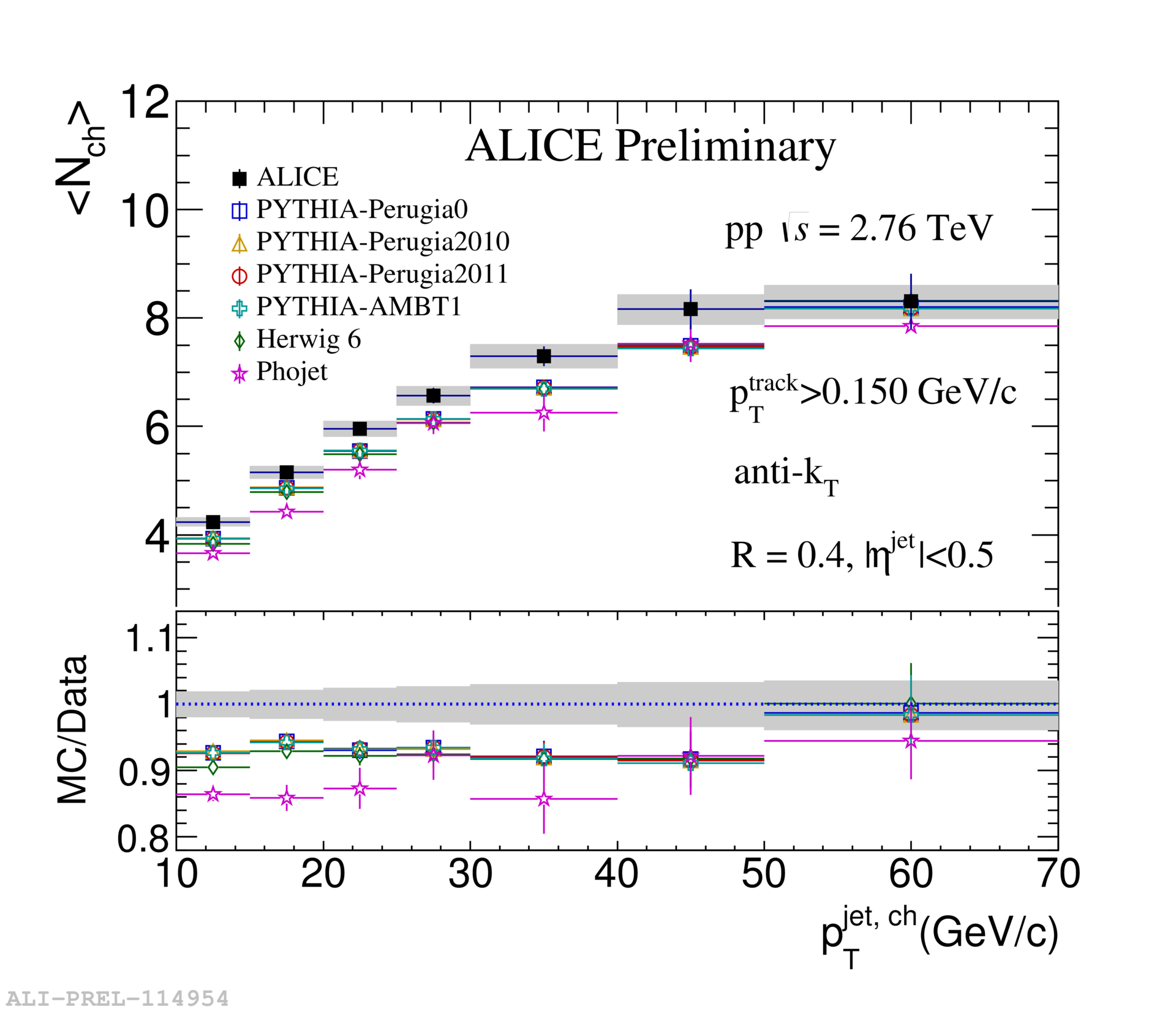}
  \includegraphics*[width = 6 cm, height = 5.6 cm]{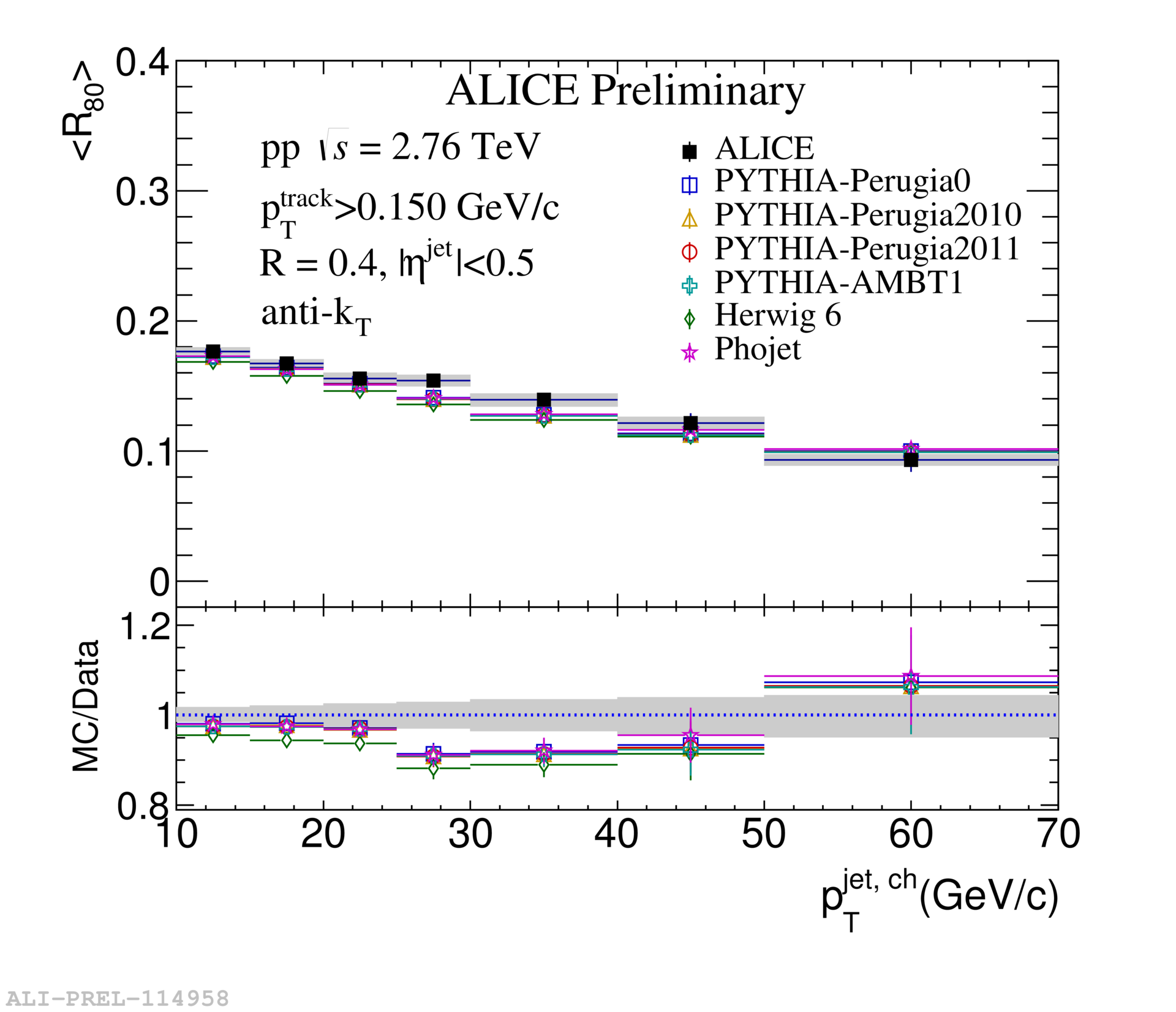}
  \caption{Top: Charged jet production cross sections for R = 0.2, 0.3, 0.4 and 0.6. Bottom left: 
  $\langle$N$_{ch}\rangle$ distributions for R = 0.4 with different MC generator predictions and
 their ratios. Bottom right: $\langle$R$_{80}\rangle$ distributions for R = 0.4 with different MC generator 
 predictions and their ratios. The vertical error bars in data and MC stand for statistical errors. The 
 boxes and the grey bands show the systematic uncertaintities.}
\label{cross}
\end{figure}
\begin{figure}[h]
  \centering
  \includegraphics*[scale=0.08]{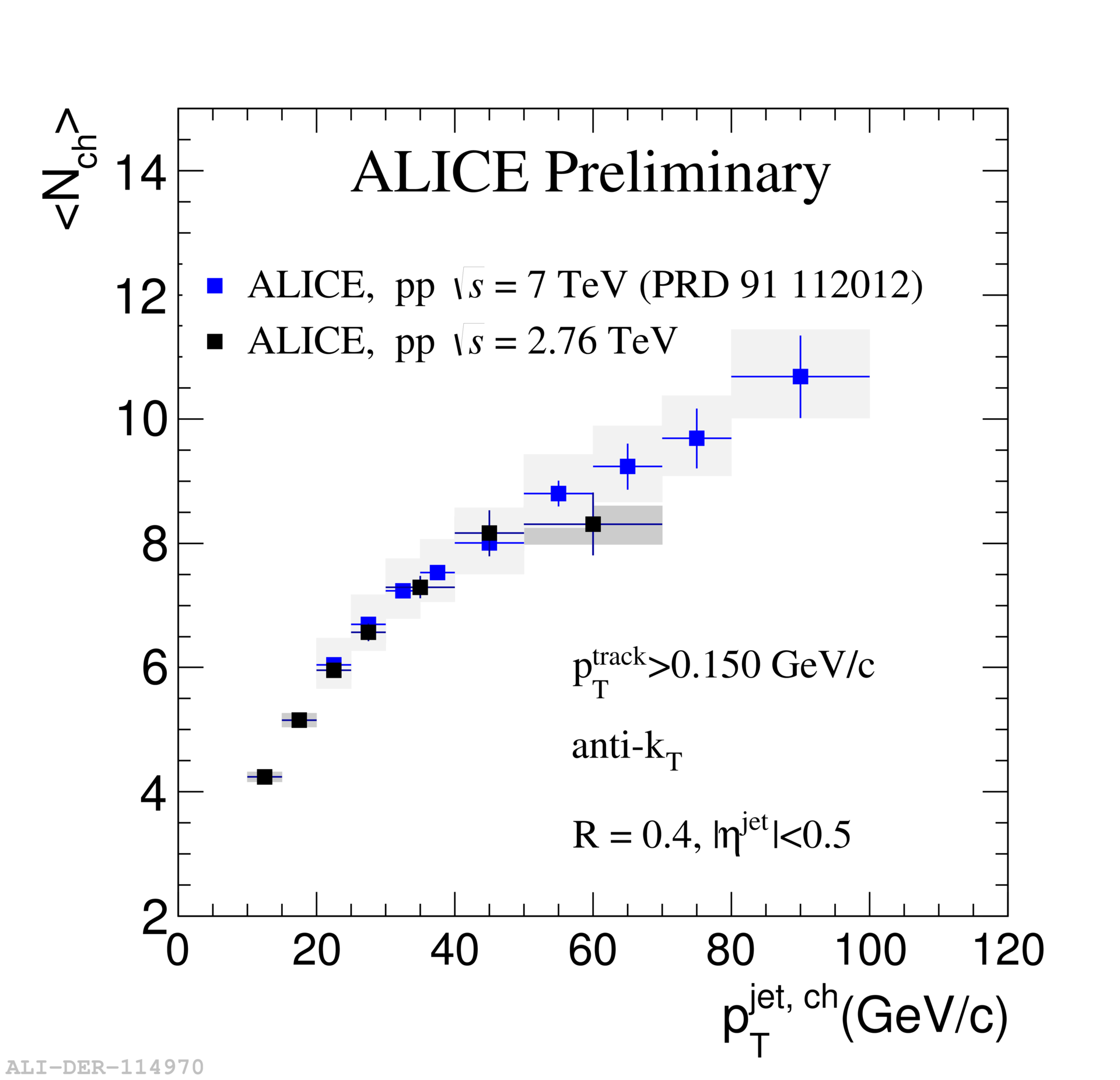}
  \includegraphics*[scale=0.08]{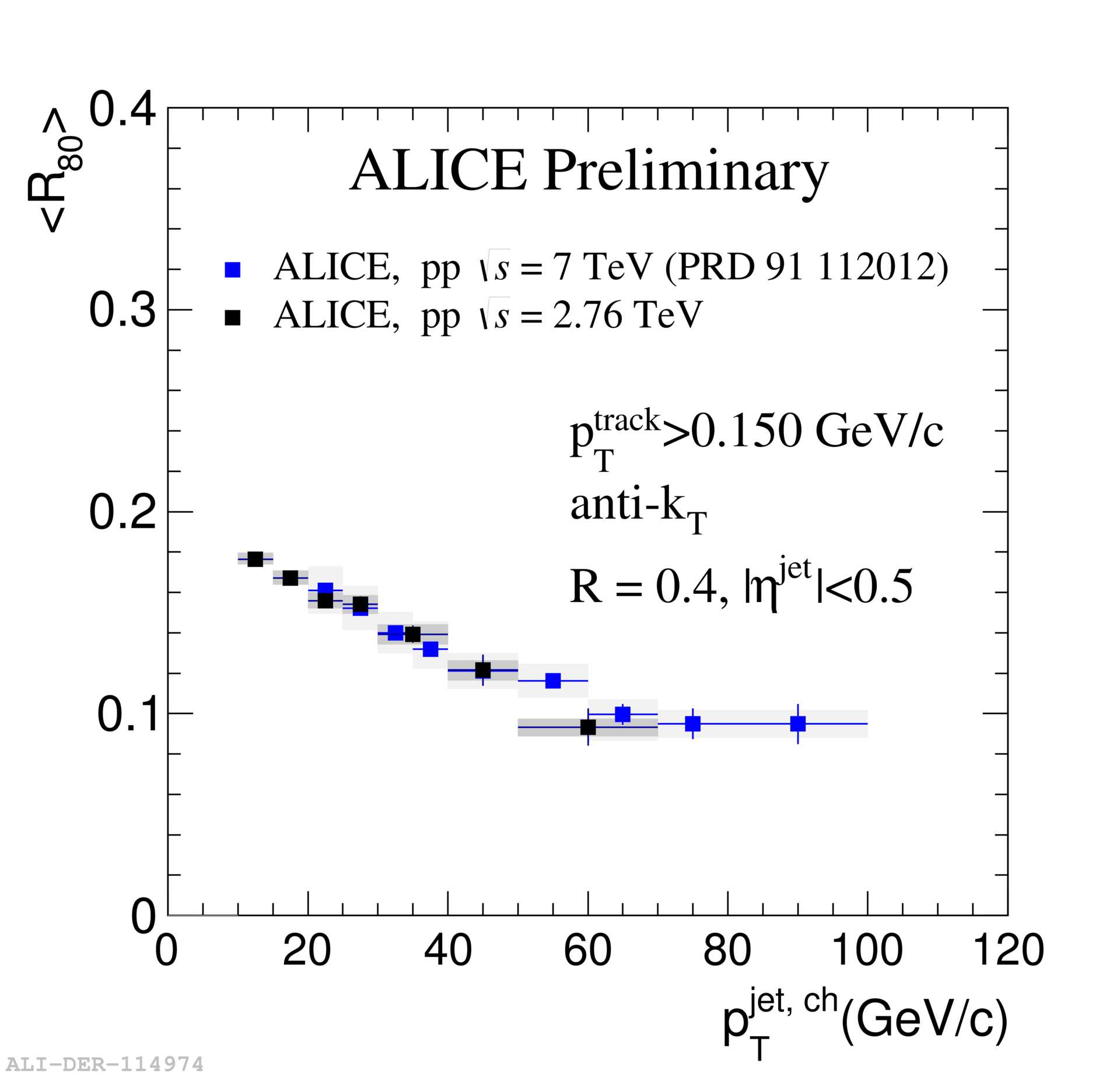}
  \includegraphics*[scale=0.08]{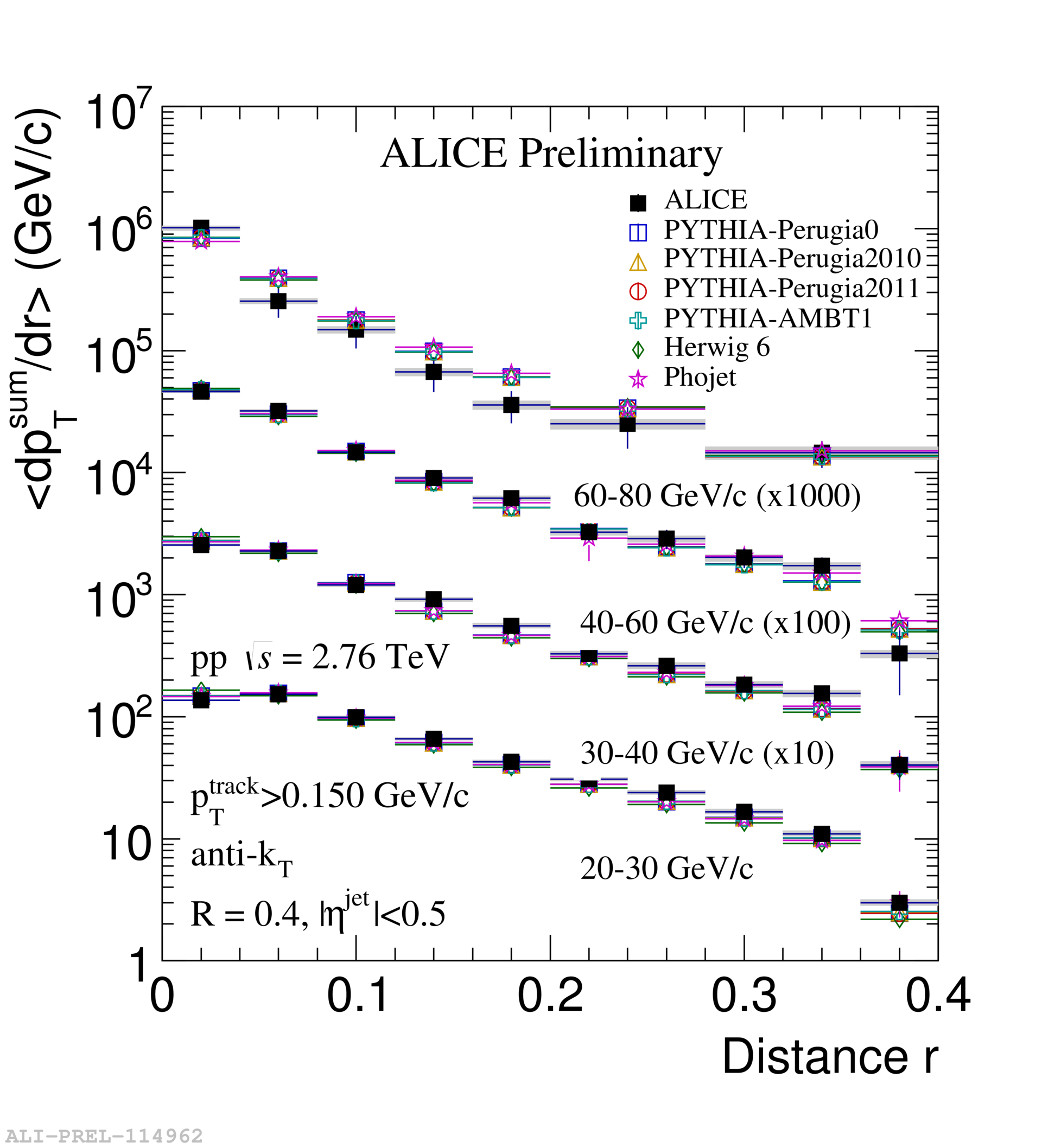}
  \includegraphics*[scale=0.07]{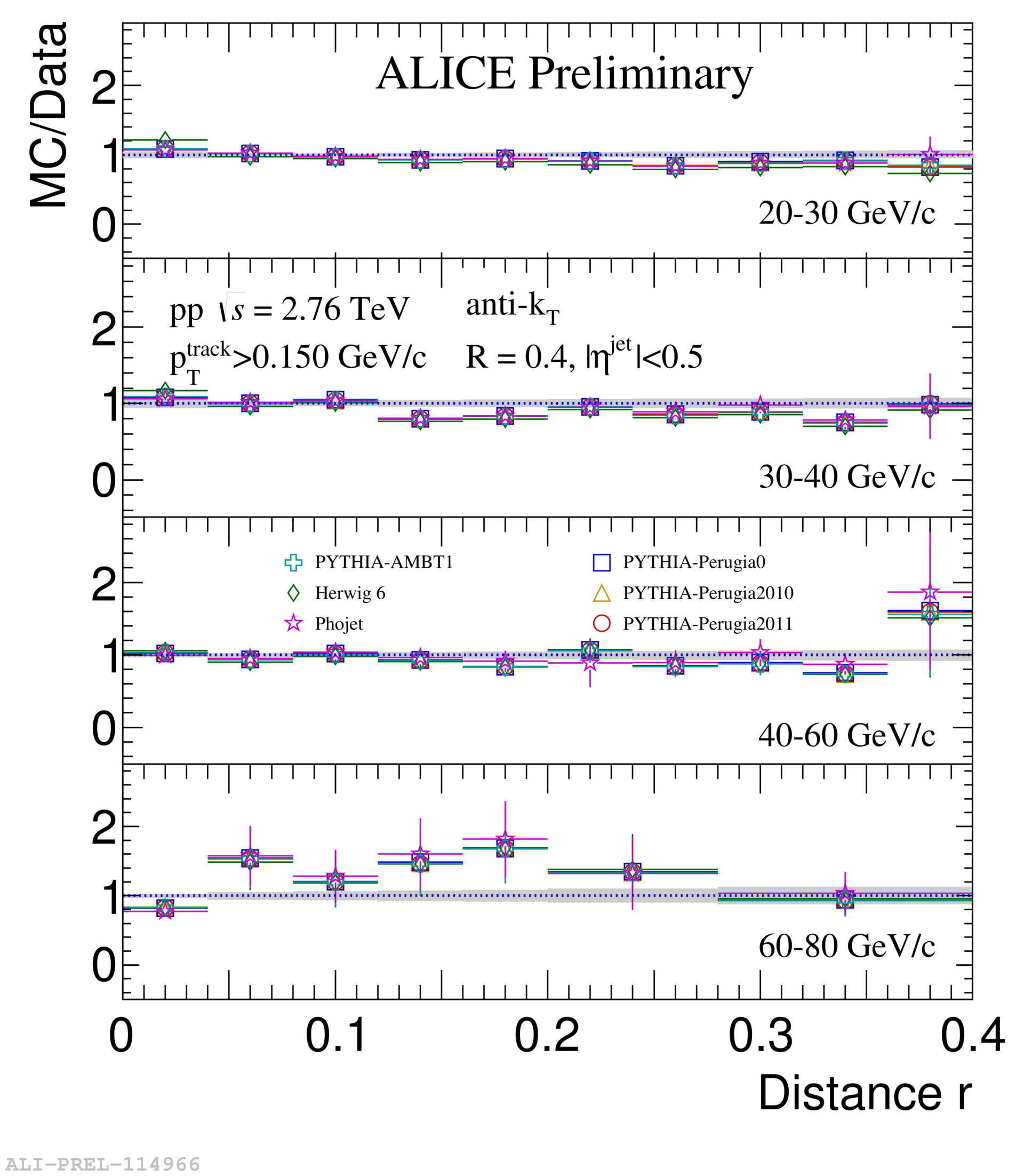}
  \caption{$\langle$N$_{ch}\rangle$ (top left) and
    $\langle$R$_{80}\rangle$ (top right) at 2.76 TeV in comparison with
    that at 7~TeV. Bottom left: p$_{T}$$^{sum}$ distributions for different jet p$_{T}$ bins for R = 0.4 with different MC 
    generator predictions. Bottom right: Ratios MC/data. The vertical error bars in data and MC stand for statistical errors. 
    The grey bands show the systematic uncertaintities.} 
  \label{nchr80comp}
\end{figure}

Figure~\ref{cross} shows jet production cross sections for R = 0.2,
0.3, 0.4 and 0.6 (top), the $\langle$N$_{ch}\rangle$ distribution for R =
0.4 (bottom left) and the $\langle$R$_{80}\rangle$ distribution for R = 0.4 (bottom right) as a function of jet p$_{T}$.
$\langle$N$_{ch}\rangle$ increases monotonically with increasing jet p$_{T}$ as seen in earlier measurements. $\langle$R$_{80}\rangle$ is contained within
half of the total radius for the lowest jet p$_{T}$,  and it decreases for higher jet p$_{T}$.
Different MC models such as HERWIG, PHOJET and different tunes of PYTHIA6 describe the data reasonably well.
The results for $\langle$N$_{ch}\rangle$ and $\langle$R$_{80}\rangle$
at 2.76~TeV are also compared to earlier measurements at 7 TeV (Fig.~\ref{nchr80comp}, top left and right) 
and as such we do not see  any $\surd s $ dependence within uncertainties.
Figure \ref{nchr80comp} (bottom left) shows radial momentum distributions for four different jet p$_{T}$ bins as a function of distance
from the jet axis. $\langle$dp$_{T}$$^{sum}$/dr$\rangle$ is largest near the jet axis and it
decreases towards the periphery. The measurements 
from various MC models are in good agreement with the data as shown in Fig~\ref{nchr80comp} (bottom right).
\section{Conclusions}
We reported measurements of charged jet cross section for R  = 0.2, 0.3, 0.4 and 0.6 and jet shapes for R = 0.4 in pp
collisions at 2.76 TeV with ALICE. $\langle$N$_{ch}\rangle$ increases with increasing
jet p$_{T}$. Measurements of $\langle$R$_{80}\rangle$ and dp$_{T}$$^{sum}$/dr reveal that high p$_{T}$ jets are more collimated.
Various MC models well reproduce the data within uncertainties except $\langle$N$_{ch}\rangle$ which agrees within 15\%. No $\surd s $ dependence is seen within the measured uncertainties.
% ---- Bibliography ----


\begin{thebibliography}{25}
\bibitem {antikt} Cacciari M, Salam G P and Soyez G, JHEP 0804 (2008) 063
\bibitem {fastjet} M. Cacciari and G. P. Salam, Phys.Lett.B 641 (2006) 57–61
\bibitem {unfold}  G. D’Agostini, Nucl.Instrum.Meth.A 362 (1995) 487–498
\bibitem {geant}  R. Brun, F. Carminati, and S. Giani, CERN-W5013, CERN-W-5013, https://cds.cern.ch/record/1082634
\bibitem {ue} ALICE Collaboration, JHEP 1207 (2012) 116
\end{thebibliography}
\end{document}